\documentclass[12pt,preprint]{aastex}

\usepackage{graphicx}
\usepackage{txfonts}
\usepackage{longtable}
\usepackage{lscape}
\usepackage{natbib}
\usepackage[usenames,dvipsnames,svgnames,table]{xcolor}
\definecolor{light-gray}{gray}{0.99}

\shorttitle{Chromospheric Dichroism in Photospheric Lines}
\shortauthors{Del Pino Alem\'an, Trujillo Bueno \& Manso Sainz}

\begin{document}

\title{Chromospheric Polarization \\in the Photospheric Solar Oxygen Infrared Triplet}

\author{{\sc Tanaus\'u Del Pino Alem\'an}\altaffilmark{1,2} \& {\sc Javier Trujillo 
Bueno}\altaffilmark{1,2,3} }
\altaffiltext{1}{Instituto de Astrof\'isica de Canarias, E-38205 La Laguna, 
Tenerife, Spain}
\altaffiltext{2}{Departamento de Astrof\'isica,
Universidad de La Laguna, E-38206 La Laguna, Tenerife, Spain}
\altaffiltext{3}{Consejo Superior de Investigaciones Cient\'ificas, Spain}

\begin{abstract}
We present multilevel radiative transfer modeling of the scattering polarization observed in the solar O {\sc i} infrared triplet around 777 nm. We demonstrate that the scattering polarization pattern observed on the solar disk forms in the chromosphere, far above the photospheric region where the bulk of the emergent intensity profiles originates. We study the sensitivity of the polarization pattern to the thermal structure of the solar atmosphere and to the presence of weak magnetic fields ($10^{-2}$ - $100$ G) through the Hanle effect, showing that the scattering polarization signals of the oxygen infrared triplet encode information on the magnetism of the solar chromosphere.        
\end{abstract}


\section{Introduction}

The absorption and scattering of anisotropic radiation in the solar atmosphere produce a richly structured linearly polarized spectrum, which is of great diagnostic potential \citep[e.g., the review by][]{Stenflo09}. In general, the physical origin of this spectrum is the radiatively induced population imbalances and quantum coherence between the magnetic sublevels of the line's levels (i.e., atomic level polarization). A weak magnetic field (i.e., such that the level's Zeeman splitting equals the inverse of its lifetime) is sufficient to produce a significant modification of the atomic level polarization (Hanle effect). In contrast to the Hanle effect, elastic collisions may destroy completely the atomic level polarization if their rates are significantly larger than the inverse of the level's lifetime. Since ground and metastable levels are relatively long-lived, they are more vulnerable than excited levels to the above-mentioned effects of magnetic fields and elastic collisions \citep[e.g.,][]{LL04}.          
 
The scattering polarization observed in a spectral line has, in general, two contributions: one due to the selective emission of polarization components, caused by the presence of atomic polarization in the upper level, and another due to selective absorption of polarization components, caused by the presence of atomic polarization in the lower level \citep{TrujilloLandi97,Trujillo99,Trujillo+02,MansoTrujillo03a}. That the atomic polarization of the lower level can contribute to the observed scattering line polarization can be easily understood by considering a one-dimensional (1D) model atmosphere, whose plasma can be unmagnetized or magnetized by a microturbulent and isotropically distributed weak magnetic field. In both cases the transfer equation for the Stokes parameter $Q$ reads (with $s$ the geometrical distance along the ray):

\begin{equation}
{d\over{ds}}Q=[\epsilon_Q\,-\,\eta_QI]\,-\,\eta_IQ,
\end{equation}
where ${\epsilon_Q}$ and $\eta_X$ (with $X=I$ or $X=Q$) are the corresponding emission and absorption coefficients. The absorption coefficient $\eta_Q$ is proportional to the alignment $\rho^2_0(\ell)$ of the lower level (which quantifies the population imbalances among the lower level sublevels), while the emission coefficient $\epsilon_Q$ is proportional to the alignment of the upper level $\rho^2_0(u)$ (which quantifies the population imbalances between the sublevels of the upper level). Since in Eq. (1) the source of Stokes $Q$ is given by the term within brackets, we have that Stokes $Q$ can be produced either because the emission coefficient $\epsilon_Q{\ne}0$ (selective emission of polarization components, which requires $\rho^2_0(u){\ne}0$) and/or because the absorption coefficient $\eta_Q{\ne}0$ (selective absorption of polarization components, which requires $\rho^2_0(\ell){\ne}0$). This last contribution is termed ``zero-field'' dichroism, because it has nothing to do with the Zeeman effect (cf., \citealt{MansoTrujillo03a}). If elastic collisions are very efficient $\epsilon_Q\,{\approx}\,\eta_Q\,{\approx}\,0$, while if they are moderately efficient we may have $\epsilon_Q\,{\ne}\,0$ and no dichroism at all (i.e., $\eta_Q\,{\approx}\,0$). In general, both $\epsilon_Q$ and $\eta_Q$ may be significant.  

\begin{figure}[t]
\begin{center}
\includegraphics[scale=0.50]{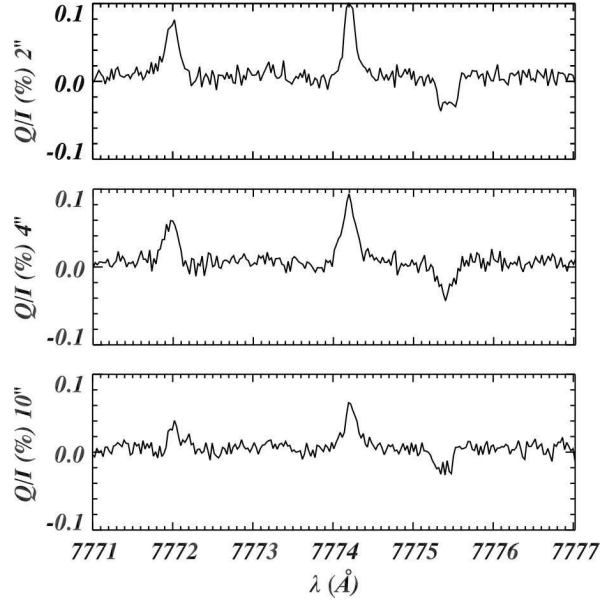}
\caption{Examples of the observations by \cite{Trujillo+01} of the scattering polarization in the IR triplet of O {\sc i} at the three indicated on-disk positions from the nearest limb. 
The positive reference direction for Stokes $Q$ is the parallel to the nearest limb.}
\label{figure-1}
\end{center}
\end{figure}

An example of solar spectral lines whose observed fractional linear polarization $Q/I$ profiles show clear signatures of such two contributions is that of the IR triplet of Ca {\sc ii} observed in quiet regions of the solar atmosphere \citep{MansoTrujillo03a}. Here elastic collisions with neutral hydrogen atoms do not destroy the polarization of the line's upper and lower levels in the chromospheric region where the core of the Stokes $I$ profiles originates. Under such circumstances the following approximate expression can be used to obtain an estimation of the amplitude of the fractional scattering polarization at the line center \citep{Trujillo03}

\begin{equation}
{{Q}\over{I}}\,{\approx}\,
{3\over{2\sqrt{2}}}(1-\mu^2)[{w^{(2)}_{J_uJ_\ell}}\sigma^2_0(J_u)-{w^{(2)}_{J_\ell J_u}}\sigma^2_0(J_\ell)],
\end{equation}
where $w^{(2)}_{JJ^{'}}$ is a coefficient introduced by \cite{Landi84} whose value depends on the total angular momentum of the line's upper and lower levels \citep[see Table 10.1 of][]{LL04} and $\sigma^2_0(J)=\rho^2_0(J)/\rho^0_0(J)$ (with $\rho^0_0(J)$ proportional to the overall population of the $J$-level) quantifies the level's fractional atomic alignment at the height in the model atmosphere where the line center optical depth is unity along the line of sight (LOS). The $\sigma^2_0(J)$ values have to be obtained by solving the statistical equilibrium equations for the multipole moments of the atomic density matrix associated to each atomic level $J$ and the Stokes-vector transfer equation for each of the radiative transitions in the atomic model.

Spectropolarimetric observations of the quiet solar atmosphere in the infrared (IR) triplet of O {\sc i} show clear scattering polarizations signals, both on the disk \citep{Trujillo+01} and off the limb \citep{SheeleyKeller03}. Figure \ref{figure-1}, adapted from a similar one in the review paper of \cite{Trujillo09}, shows some close to the limb examples of the {\em on-disk} observations by \cite{Trujillo+01}. Note that while the lines at 7772 \AA\ (hereafter, line 1) and 7774 \AA\ (hereafter, line 2) showed positive $Q/I$ signals, with line 2 slightly more polarized than line 1, negative $Q/I$ values were found for the 7775 \AA\ line (hereafter, line 3). Given the complexity of the radiative transfer problem, \cite{BelluzziTrujillo11} assumed a plane-parallel layer of O {\sc i} atoms illuminated by the solar continuum radiation field, and applied Eq. (2) after solving the statistical equilibrium equations for the $\sigma^2_0(J_i)$ values \citep[see also][]{Trujillo09}. These authors concluded that the observed $Q/I$ signals can be qualitatively explained if there is a sizable  amount of atomic polarization in the upper and lower line levels around the height in the solar photosphere where the optical depth at the line center is unity along the line of sight, with zero-field dichroism playing a significant role.           

However, as shown in this letter, the situation is more subtle and interesting. It turns out that in the solar photosphere elastic collisions between the O {\sc i} and H {\sc i} atoms destroy the atomic level polarization of the O {\sc i} IR triplet (i.e., $\sigma^2_0(J_\ell)\,{\approx}\,\sigma^2_0(J_u)\,{\approx}\,0$ in the photospheric region where the bulk of the emergent intensity profiles originates, both for the lower and upper levels). Therefore, according to the approximate Eq. (2) the O {\sc i} IR triplet should show negligible polarization, contrary to the on-disk observations of Fig. 1. In order to find the true physical origin of the observed scattering polarization, we solved numerically the full (non-local and non-linear) radiative transfer problem iteratively and calculated the emergent Stokes $I$ and $Q$ profiles using the resulting self-consistent values of the emission and absorption coefficients at each point along the line of sight. The main result we want to highlight here is that the scattering polarization signals observed in the oxygen IR triplet are produced in the solar chromosphere. As we shall see below, their physical origin is the selective emission and selective absorption (dichroism) processes caused by the presence of atomic level polarization within the solar chromosphere.      

\section{Theoretical approach}

Our theoretical approach is based on the quantum-mechanical density matrix theory explained in \cite{LL04}, which assumes that the absorption and re-emission of photons are statistically independent events. \cite{BelluzziTrujillo11} have investigated the impact of quantum interference between pairs of sublevels pertaining to different fine-structure $J$-levels (hereafter, $J$-state interference), showing that it can be neglected for modeling on-disk observations of the scattering polarization in the O {\sc i} IR triplet. For the case of a multilevel atom neglecting $J$-state interference the above-mentioned theory is a reliable approach whenever the radiation field that pumps the atomic system is flat over a frequency interval larger than the natural width of the levels. This is a suitable assumption for modeling the oxygen IR triplet and we have developed a radiative transfer code that solves jointly the statistical equilibrium equations for the multipole moments  of the atomic density matrix associated to each atomic level $J$ and the Stokes-vector transfer equation for each of the radiative transitions in the atomic model. The method of solution is based on a Jacobian multilevel iterative scheme \citep{Trujillo99,MansoTrujillo03b} and on the BESSER formal solver of the Stokes-vector transfer equation proposed by \cite{StepanTrujillo13}, which we have used also for computing the emergent Stokes $I$ and $Q$ profiles that give the $Q/I$ ones shown below. 

The atomic model, which has 23 O {\sc i} $J$-levels and the ground level of O {\sc ii}, is an improved version of the model atom used by \cite{CarlssonJudge93}. The model takes into account 43 bound-bound radiative transitions, with Einstein coefficients taken from the NIST database (\citealt{NIST}). The number density of neutral oxygen atoms at each atmospheric height was calculated taking into account the ionizations from each O {\sc i} level, with the cross sections taken from the Opacity Project\footnote{http://vizier.u- strasbg.fr/topbase/topbase.html}. The bound-bound inelastic collisions are calculated using the approximation of \cite{VanRegemorter62} if the corresponding transition is radiatively allowed and with the approximation of \cite{BelyVanRegemorter70} if the corresponding transition is radiatively forbidden. The bound-free inelastic collisions are calculated following the formulae given by \cite{Cox00}. Collisions with protons are taken into account for the ground term levels, following \cite{Haisch+77}. 

The lines of the O {\sc i} IR triplet share the same (metastable) lower level, whose angular momentum is $J_\ell=2$; the angular momentum values of the three upper levels of increasing energy are 1, 2 and 3. Therefore all these levels can carry atomic alignment and contribute to the observed scattering polarization. When observed on the solar disk at a line of sight with $\mu={\rm cos}{\theta}=0.1$ ($\theta$ being the heliocentric angle), the mean formation heights of the line core intensity profiles are located around 200 km above the solar surface.  At such low photospheric heights the number density of neutral hydrogen atoms is of the order of $10^{16}{\rm cm}^{-3}$, so we have to investigate if elastic collisions between neutral oxygen and hydrogen atoms are able to depolarize the O {\sc i} IR triplet levels. To this end, we have used equation (7.108) of \cite{LL04}, which is suitable to estimate the rates ${\rm D}_i^{(2)}$ of such elastic collisions 
for the lower ($i=\ell$) and upper ($i=u$) levels of the O {\sc i} IR triplet.

\begin{figure}[t]
\begin{center}
\includegraphics[scale=0.85]{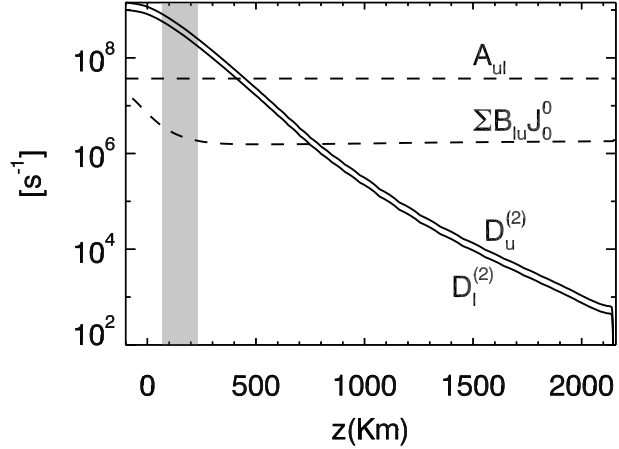}
\caption{The variation with height in a 1D semi-empirical model of the quiet solar atmosphere of the rates of elastic collisions with H {\sc i} atoms, for the levels of the O {\sc i} IR triplet (solid curves). The upper and lower dashed lines  indicate the height variation of the rates of spontaneous radiative emissions and of radiative absorptions, respectively. The shaded area indicates the atmospheric heights where the optical depth at the center of the O {\sc i} lines is unity for line of sights with $0.1\,{\le}\,{\mu}\,{\le}\,1$.} 
\label{figure-2}
\end{center}
\end{figure}

Figure \ref{figure-2} shows the variation with height in a semi-empirical model of the quiet solar atmosphere of the rates of elastic collisions with neutral hydrogen, for the lower level and the three upper levels of the O {\sc i} triplet. These collisional rates have to be compared with the rates of spontaneous radiative emissions (upper dashed line) and with the rates of radiative absorptions (lower dashed curve). As can be seen, below 400 km both radiative rates are smaller than the rates of elastic collisions and, therefore, we may expect that in the low photosphere of the quiet Sun (see the  shaded region in Fig. \ref{figure-2}) elastic collisions with neutral hydrogen atoms are efficient in destroying the atomic alignment that anisotropic radiation pumping processes induce in the O {\sc i} IR triplet levels. 

\begin{figure}[t]
\begin{center}
\includegraphics[scale=0.60]{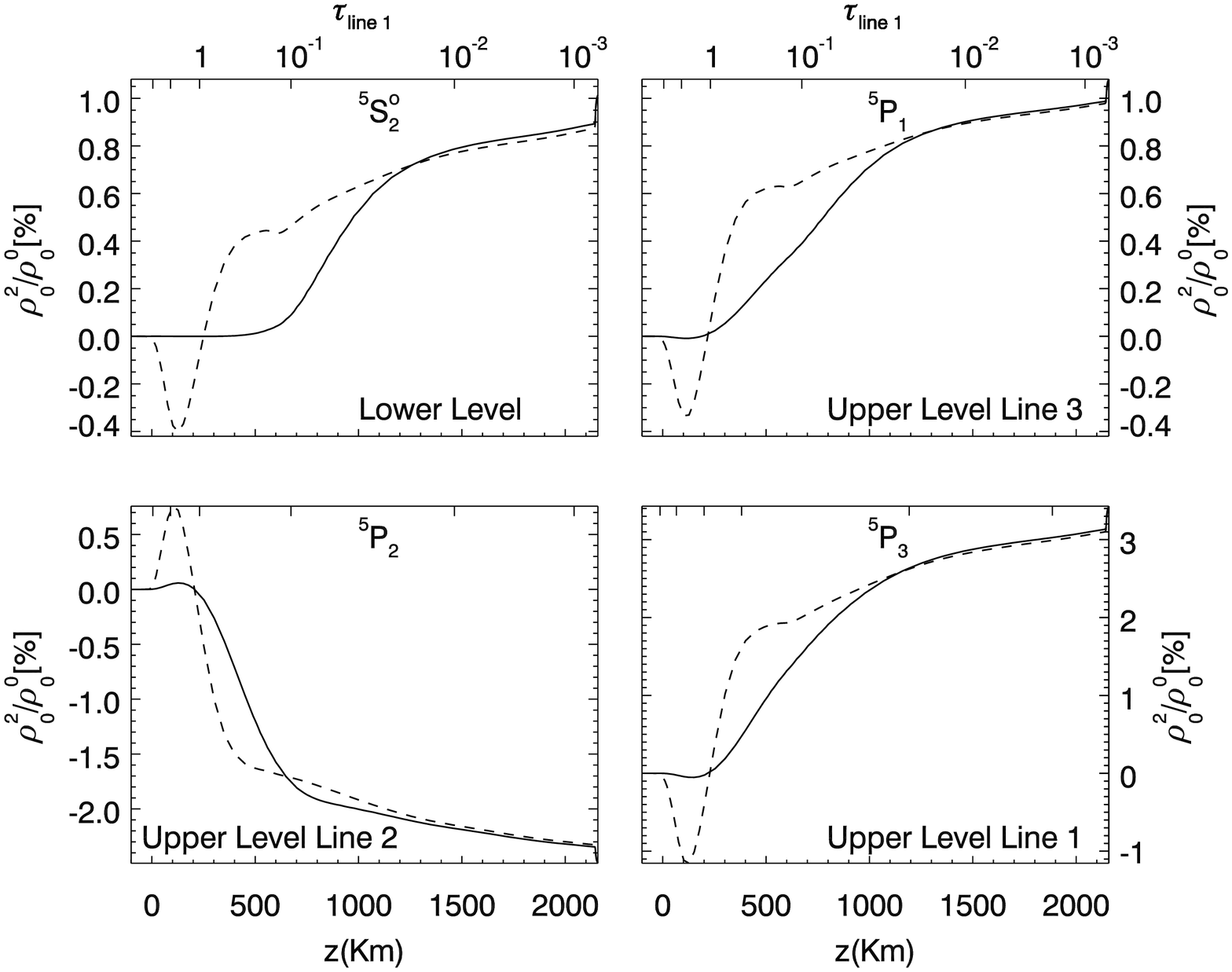}
\caption{The height variation of the fractional atomic alignment of the O {\sc i} IR triplet levels calculated in the FAL-C semi-empirical model. The solid curves take into account the elastic collisions with neutral hydrogen atoms while the dashed curves neglect them. Note that the top horizontal axes give the line center optical depth of line 1 along a LOS with $\mu=0.1$.}
\label{figure-3}
\end{center}
\end{figure}

\section{Results}

This section shows and discusses the results we have obtained by solving jointly the statistical equilibrium and radiative transfer equations, using the above-mentioned multilevel oxygen model and semi-empirical models of the quiet solar atmosphere. 

\begin{figure}[htb]
\begin{center}
\includegraphics[scale=0.62]{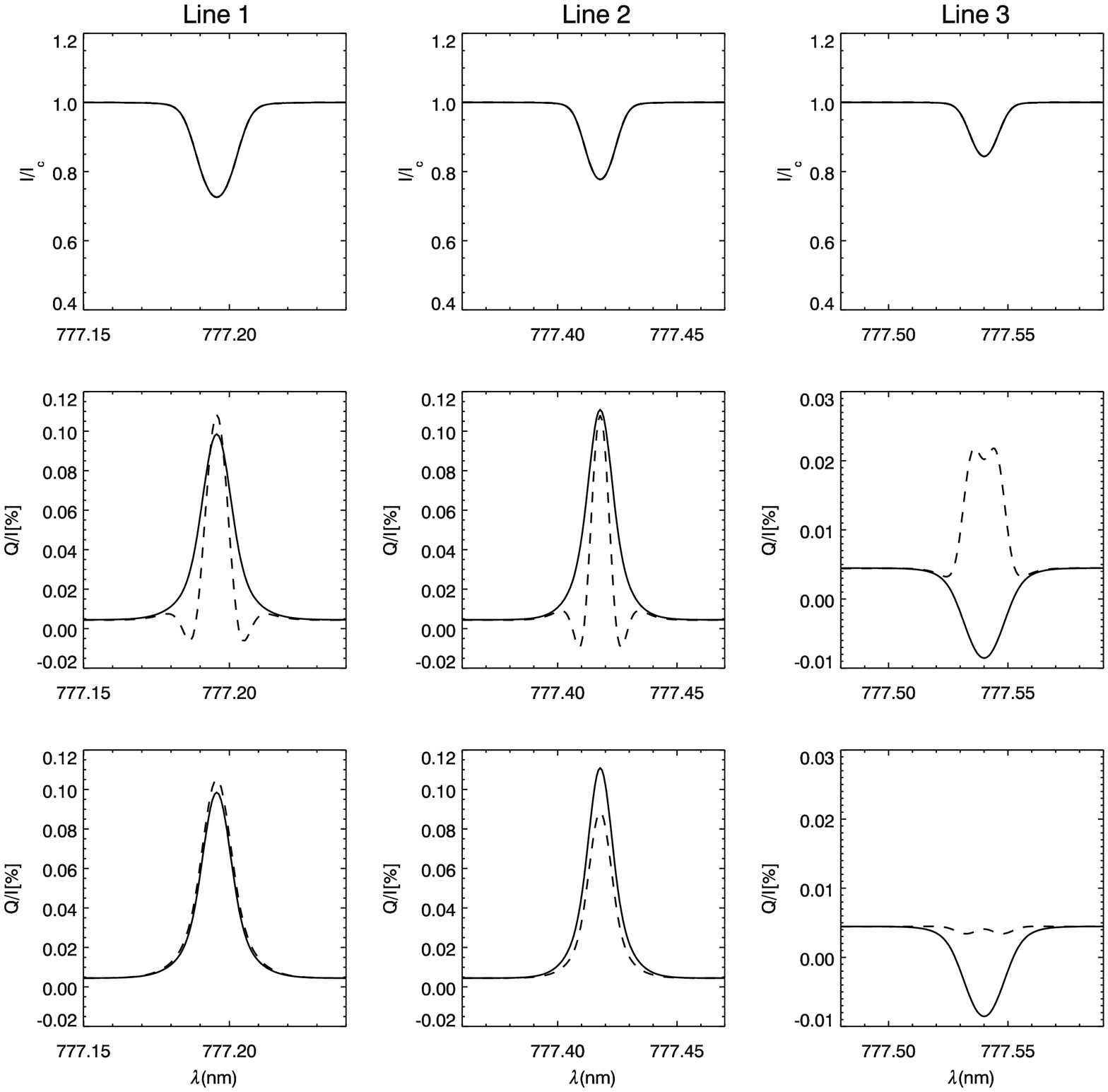}
\caption{The Stokes $I$ and $Q/I$ profiles of the IR triplet of O {\sc i} calculated in the FAL-C semi-empirical model for a line of sight with $\mu=0.1$. The positive reference direction for Stokes $Q$ is the parallel to the nearest limb. Top panels: the Stokes $I$ profiles normalized to the continuum intensity. Central panels: $Q/I$ profiles calculated with (solid curves) and without (dashed curves) elastic collisions. Bottom panels: $Q/I$ profiles neglecting the contribution of dichroism (dashed curves) and taking it into account (solid curves). 
}
\label{figure-4}
\end{center}
\end{figure}

\subsection{The atomic polarization and the emergent spectral line polarization}

Figure \ref{figure-3} shows the height variation of the self-consistent fractional atomic alignment values of the O {\sc i} IR triplet levels. The solid curves take into account the impact of elastic collisions with neutral hydrogen atoms, while the dashed curves correspond to the case in which elastic collisions are neglected. To obtain such $\sigma^2_0(J)=\rho^2_0(J)/\rho^0_0(J)$ values we numerically solved the ensuing radiative transfer problem of scattering line polarization in the quiet Sun model C of \cite{Fontenla+93} (hereafter, FAL-C model) using the above-mentioned realistic multilevel oxygen model and taking into account the impact of the hydrogen Lyman-$\beta$ pumping on the populations of the oxygen levels (see \citealt{MillerRicciUitenbroek02}). As shown by the solid curves of the figure, in the model's photosphere elastic collisions strongly reduce the polarization of the O {\sc i} IR triplet levels, yielding $\sigma^2_0(J)\,{\approx}\,0$ in the photospheric formation region of the Stokes-$I$ profiles. Therefore, since this implies that $\epsilon_Q{\approx}0$ and $\eta_Q{\approx}0$ in the solar photosphere, the observed $Q/I$ signals shown in Fig. \ref{figure-1} cannot be produced in the same atmospheric region (the photosphere) where the bulk of the Stokes $I$ profiles originates.

Interestingly, if elastic collisions are fully neglected (see the dashed curves of Fig. \ref{figure-3}) the ensuing $Q/I$ profiles of the calculated emergent radiation do not agree with the observations of Fig. \ref{figure-1} (see the dashed curves of the central panels of Fig. \ref{figure-4} and note that without elastic collisions the calculated $Q/I$ signal for line 3 is positive, while the observed one is negative). However, as shown by the solid curves of the central panels of Fig. \ref{figure-4}, the calculated $Q/I$ profiles qualitatively agree with the observed ones as soon as the impact of the elastic collisions is taken into account.

In retrospect, we can clarify the reason why the application of the approximate Eddington-Barbier Eq. (2) used by \cite{Trujillo09} and \cite{BelluzziTrujillo11} allowed them to explain qualitatively the observations of Fig. 1, although missing the interesting point explained in this Letter (i.e., that the observed scattering polarization originates in the chromosphere). As mentioned in Section 1, those authors considered a plane-parallel layer of O {\sc i} atoms and solved the statistical equilibrium equations for obtaining the $\sigma^2_0(J)$ values assuming that the anisotropic pumping is exclusively due to the continuum radiation coming from the underlying solar visible surface. Such a predominantly vertical incident radiation is characterized by a positive anisotropy factor, which in the absence of elastic collisions leads to significant $\sigma^2_0(J)$ values with signs similar to those found in Fig. 3 at chromospheric heights.      

\subsection{Zero-field chromospheric dichroism}
   
In order to demonstrate that zero-field dichroism within the solar chromosphere produces clear observational signatures, we have calculated the $Q/I$ profiles that result when the contribution of the $\eta_QI$ term of Eq. (1) is neglected above 500 km in the FAL-C semi-empirical model. The dashed curves of the bottom panels of Fig. \ref{figure-4} show the ensuing results, along with the previously shown $Q/I$ profiles (solid curves of Fig. \ref{figure-4}) that take into account the full contribution. Clearly, the $Q/I$ profile of line 3 is dominated by zero-field chromospheric dichroism. The reason why the $Q/I$ signal of line 3 is dominated by the $\eta_QI$ term of Eq. (1) (i.e., by zero-field dichroism) is easy to understand by noting that $\sigma^2_0(l){\approx}\sigma^2_0(u)$ for line 3 (see Fig. \ref{figure-3}), while ${w^{(2)}_{J_\ell J_u}}{\approx}0.6$ and ${w^{(2)}_{J_u J_\ell}}=0.1$. Moreover, zero-field dichroism has a significant impact on the fractional polarization line-center amplitudes of lines 1 and 2. As seen in the bottom panels of Fig. \ref{figure-4}, only when the full impact of zero-field dichroism is taken into account we obtain that line 1 is slightly less polarized than line 2, in agreement with the observations of Fig. \ref{figure-1}.

 \subsection{Sensitivity to the model's thermal structure}
 
The scattering polarization depends on the anisotropy of the radiation field that produces optical pumping within the solar atmosphere, which in turn depends on the thermal structure of the atmospheric plasma among other factors. For this reason, we have investigated also the sensitivity of the emergent $Q/I$ profiles of the O {\sc i} IR triplet to the thermal structure of the model atmosphere. To that end, we carried out radiative transfer calculations in two semi-empirical models: the FAL-C model already considered and the ${\rm M_{CO}}$ model of \cite{Avrett95} which is significantly cooler around a height of 1000 km (e.g., see figure \ref{figure-2} of \citealt{MansoTrujillo10}). 

Since the two atmospheric models are practically identical below 500 km, we point out that the emergent Stokes-$I$ profiles are similar. However, the calculated $Q/I$ profiles turn out to be very different, in agreement with the previously reported point that the scattering polarization signals are produced within the chromosphere where the two models clearly differ. Interestingly enough, only the $Q/I$ profiles calculated in the (hotter) FAL-C model agree qualitatively with the observations of Fig. \ref{figure-1} (e.g., we find that in the ${\rm M_{CO}}$ model line 3 has a positive signal, contrary to the observed $Q/I$ profile). This additional result emphasizes that the observed scattering polarization is also of interest for constraining the thermal structure of the solar atmosphere. 

\subsection{Hanle sensitivity to the presence of a magnetic field}
 
The critical magnetic field $B_{\rm H}$ (in gauss) needed to produce a significant modification of the polarization of an atomic level via the Hanle effect is that for which the level's Zeeman splitting (in frequency units) equals the inverse of its radiative lifetime: $B_{\rm H}=(1.137{\times}10^{-7})/(g\,t_{\rm life})$ (with $g$ the level's Land\'e factor and $t_{\rm life}$ its radiative lifetime in seconds). Since the lower level of the O {\sc i} IR triplet is metastable (i.e., long-lived), milligauss fields are sufficient to significantly alter its atomic polarization, while the critical fields of the three upper levels of the O {\sc i} IR triplet are 1.7 G, 2.2 G and 2.5 G for the upper levels of lines 3, 2 and 1, respectively.

\begin{figure}[t]
\begin{center}
\includegraphics[scale=0.65]{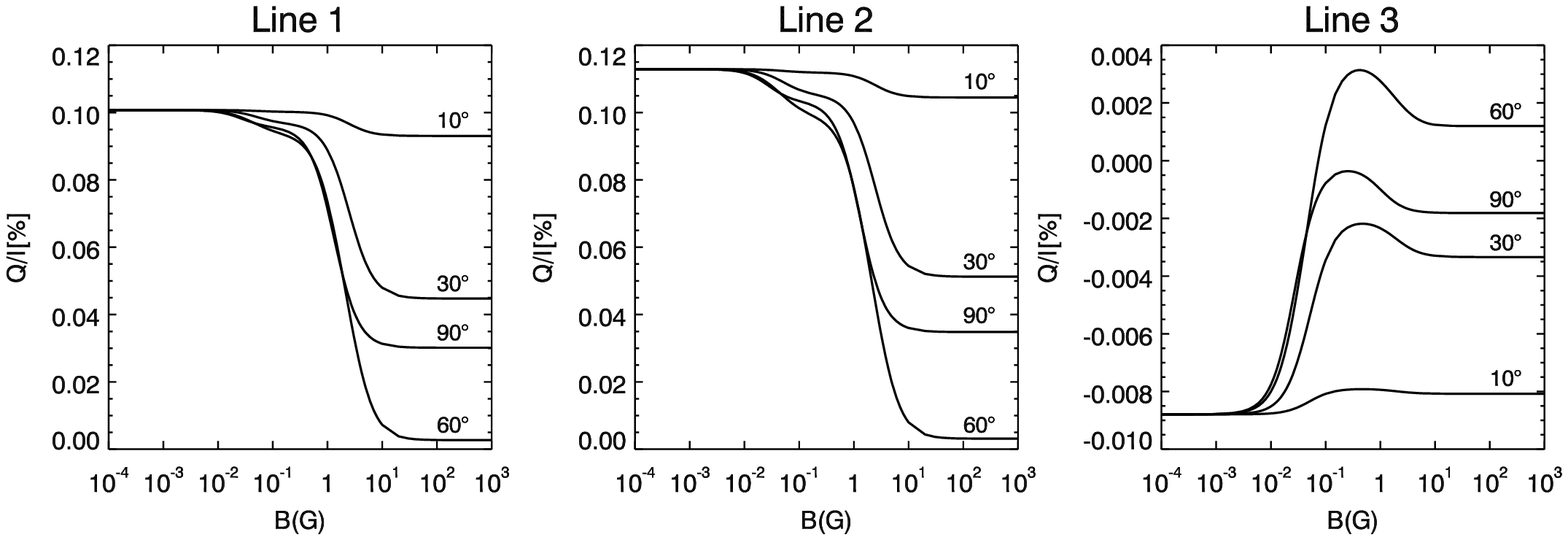}
\caption{The $Q/I$ line-center amplitudes of the IR triplet of O {\sc i} versus the strength of a micro-structured magnetic field with random azimuth and the indicated inclination with respect to the local vertical, calculated in the FAL-C semi-empirical model for a line of sight with $\mu=0.1$. The positive reference direction for Stokes $Q$ is the parallel to the nearest limb.}
\label{figure-7}
\end{center}
\end{figure}

Figure \ref{figure-7} illustrates the sensitivity to the Hanle effect of the scattering polarization signal at the center of the O {\sc i} IR triplet lines, assuming random-azimuth magnetic fields of fixed inclination. As shown in the figure, the scattering polarization of these O {\sc i} lines react to magnetic fields between $10^{-2}$ and 100 gauss. Finally, we point out that the response functions of the line-center amplitudes of the $Q/I$ profiles to magnetic strength perturbations in the FAL-C semi-empirical model show significant values between 600 and 1600 km, approximately, with their peaks around 1000 km.         
 
\section{Concluding comments}

The intensity profiles of the solar O {\sc i} IR triplet originate mainly in the low photosphere. However, we have shown that their scattering polarization signals are generated above the photosphere, where elastic collisions with neutral hydrogen atoms do not destroy the atomic polarization that anisotropic radiation pumping processes induce in the oxygen levels. The physical origin of the observed linear polarization signals is the selective absorption (dichroism) and selective emission of polarization components caused by the atomic level polarization of the chromosphere. The scattering polarization of the O {\sc i} IR triplet is sensitive to the thermal and magnetic structure of the chromosphere. In particular, via the Hanle effect the scattering polarization amplitudes are sensitive to magnetic fields with strengths between milligauss and a few gauss, and with response functions to magnetic field perturbations that peak around 1000 km above the model's visible surface. 

New spectropolarimetric observations of the O {\sc i} IR triplet would be very welcomed. Of special interest would be to observe regions with different levels of magnetic activity (e.g., within and outside coronal holes) and to pay attention to the sign of the scattering polarization signals in the three lines and to their relative polarization amplitudes. By contrasting such observables with radiative transfer calculations in increasingly realistic models of the solar atmosphere we may hope to learn more about the thermal and magnetic structure of its enigmatic chromosphere.      

\acknowledgements
We are grateful to Rafael Manso Sainz (IAC) for helpful discussions and suggestions.  
Financial support by the Spanish Ministry of Economy and Competitiveness through projects AYA2010-18029, AYA2014-60476-P and CONSOLIDER INGENIO CSD2009-00038 are gratefully acknowledged.

\end{document}